%% file: snapstab.arxiv.tex
\documentclass[smallextended]{svjour3}
\usepackage[pagewise]{lineno}

\usepackage{palatino}
\usepackage[utf8]{inputenc}
\usepackage[T1]{fontenc}

\usepackage[alwaysadjust]{paralist} %
\usepackage{multicol}

\usepackage{amsmath,amssymb} %
\newenvironment{qedproof}[1][]{\begin{proof}#1}{\flushright$\square$\end{proof}}

\usepackage{xspace} %
\newcommand{\ie}{\textit{i.e.}\xspace}
\newcommand{\macro}[2]{ \providecommand{#1}{{\ensuremath{#2}}\xspace}}

\macro{\N}{\mathbb N}
\macro{\Q}{\mathbb Q}
\macro{\R}{\mathbb R}
\macro{\Z}{\mathbb Z}
\macro{\bD}{\bf D}
\macro{\bG}{\bf G}
\macro{\bH}{\bf H}
\macro{\bK}{\bf K}
\macro{\dfam}{\mathcal F}
\macro{\out}{\textsc{Out}}
\macro{\mem}{\textsc{Mem}}

\macro{\unaryminus}{\scalebox{0.55}[1.0]{\( - \)}}
\usepackage[vlined,linesnumbered,ruled]{algorithm2e}
\usepackage{tikz} %

\macro{\bG}{\mathbf G}
\macro{\allg}{\mathcal G} %
\macro{\allanon}{\allg_\mathrm{a}} %
\macro{\allid}{\allg_\mathrm{id}} %
\macro{\gfam}{\mathcal F}
\macro{\size}{\mathrm{size}}
\macro{\carto}{\textsc{Carto}}
\macro{\algo}{\textsc{Algo}}
\macro{\algo}{\mathbb P}
\macro{\spec}{\mathcal S}
\macro{\causal}{\mathcal C}

\newcounter{theoenumcounter}
\def\theoenumlabel{\thetheorem.\roman{theoenumcounter}}

\newenvironment{theoenum}{\begin{list}{\theoenumlabel}{%
      \usecounter{theoenumcounter}      
}      %
  }{\end{list}}

\newcounter{rrulec}
\newenvironment{rrule}[1]{%
  \begin{list}{\therrulec~:}
    { 
      \usecounter{rrulec}
      \newcommand{\therrule}{#1}
    }
  }
  {\end{list}}
\newcommand{\ritem}[4][\therrule\arabic{rrulec}]{
  \renewcommand{\therrulec}{#1}
  \pagebreak[3]
  \item {\bf #2 }\nopagebreak[4] 
    \begin{compactitem}
    \item[] \underline{\em Guard~:}
      \begin{compactitem}
        #3
      \end{compactitem}
    \item[] \underline{\em Action~:}
      \begin{compactitem}
        #4
      \end{compactitem}
    \end{compactitem}{\vskip 0.5cm}
}

\usepackage{fancyhdr}
\fancypagestyle{plain}{%
  \fancyhf{} %
  \fancyfoot[R]{{{\thepage}}} %
}
\fancypagestyle{empty}{%
   \fancyhf{} %

 }

\begin{document}
\title{Snap-Stabilizing Tasks in Anonymous Networks}
\author{Emmanuel Godard%
}
\institute{Aix-Marseille Université, CNRS, LIS, Marseille, France}

\pagestyle{plain}
\maketitle

\begin{abstract}
  We consider snap-stabilizing algorithms in anonymous networks.
  Self-stabilizing algorithms are well known fault tolerant algorithms : a
  self-stabilizing algorithm will eventually recover from arbitrary
  transient faults.
  On the other hand, an algorithm is snap-stabilizing if it can withstand
  arbitrary initial values and immediately satisfy its safety requirement.
  It is a subset of self-stabilizing algorithms.
  Distributed tasks that are solvable with self-stabilizing algorithms in anonymous networks
  have already been characterized by Boldi and Vigna in \cite{BVselfstab}.

  In this paper, we show how the more demanding snap-stabilizing algorithms can be handled
  with standard tools for (not stabilizing) algorithms in anonymous
  networks. We give a characterization of which tasks are
  solvable by snap-stabilizing algorithms in anonymous networks.
  We also present a snap-stabilizing version of Mazur\-kie\-wicz' enumeration algorithm. 

  This work exposes, from a
  task-equivalence point of view, the complete correspondence in anonymous networks between self or
  snap-stabilizing tasks and distributed tasks with various termination
  detection requirements.
\end{abstract}

\setcounter{page}{1}
\setcounter{footnote}{0}

\section{Introduction}

In the world of fault-tolerance, distributed tasks that admits self-stabilizing solutions have been long
studied \cite{dolev}.
An algorithm is self-stabilizing if, starting from arbitrary initial values
in the registers used by the algorithm, it can eventually stabilize to
a correct final value. In particular, when looking at some computed
values, the algorithm can output incorrect values as long as it
eventually outputs correct ones.

In contrast, an algorithm is snap-stabilizing if it can withstand arbitrary
initial values and output only correct values \cite{snapstab}. 
Snap-stabilizing tasks form a subset of self-stabilizing tasks where
the algorithm is required to retain computed values until it is
''sure'' that they 
are correct.
Snap-stabilizing algorithms have really interesting properties, they
can withstand arbitrary transient failures, while at the same time,
improving on self-stabilizing algorithms about a key point : the
stabilization moment is not unknown : when a response is given, it is
correct.

We present here the first characterization of snap-stabilizing tasks
on anonymous networks. Not only we are reusing techniques borrowed
from the study of the non-stabilizing tasks in anonymous networks and
show they apply also here, but we complete the correspondence between
self/snap-stabilizing tasks and termination detection.

How does snap-stabilizing tasks differ from self-stabilizing tasks has
not been considered so far in anonymous networks
to the best of our knowledge.
Here we show that, on anonymous networks,
there are tasks that admit self-stabilizing solutions but that have no
snap-stabilizing ones.
We show that the difference between self and snap stabilization is
actually the same one gets with non-stabilizing tasks when considering
implicit vs explicit termination.
This result completes the understanding of the computability power
of fault-tolerant and non fault-tolerant algorithms.

\subsection{Our Result}
We give the first characterization of the computability of snap-stabilization.
In order to show that it complements known results about
self-stabilizing and non self-stabilizing tasks in anonymous
networks, we recall the previous equivalence established by Boldi and
Vigna.
Solving a task means solving a given specification linking inputs labels to output labels
for a given set of
graphs. Informally an algorithm has implicit termination if it is allowed
to write numerous times a (tentative) solution in the dedicated \out
register. An algorithm has explicit termination when it is possible to
write in \out only once. Whenever the \out register is defined,
this means that (locally) the algorithm has terminated  its computation.

\begin{theorem}[Boldi and Vigna \cite{BVanonymous,BVselfstab}]
A task is solvable on a family of anonymous networks by a
self-stabilizing algorithm if and only if it is 
solvable with implicit termination.
\end{theorem}

The ``only if'' part being obvious, the merit of \cite{BVselfstab} is
to show that there is a universal algorithm to solve tasks (that are
at all solvable) by a self-stabilizing algorithm on anonymous
networks, and that the condition for solvability (informally speaking:
stability of the specification by lifting) is exactly the one
required by implicit termination.  In other words, once a task is
solvable with implicit termination, it admits a reliable
self-stabilizing solution without any additional condition.

\begin{theorem}[this paper]\label{snapequiv}
A task is solvable on a family of anonymous networks by a
snap-stabilizing algorithm if and only if it is 
solvable with explicit termination.
\end{theorem}

As in the Boldi and Vigna result, the ``only if'' part is immediate. We
therefore focus on establishing the ``if'' part. So the main
contribution of this paper is a universal snap-stabilizing algorithm that solves
the task at hand if this task satisfies the condition for being
solvable by an algorithm with explicit termination. 

This condition is given in Theorem~\ref{CN}. It is the same as
the one given in \cite{CGMterm} for solvability with explicit termination.
We first prove our results for terminating tasks in the
asynchronous model, then we show how to extend the technique for long
lived tasks in the synchronous model (for simplicity of exposition).

The roadmap is the following.  Section 2 introduces the model of
computation and the definition of snap-stabilizing algorithms. Section
3 introduces the algebraic tools that are necessary to express the
condition in Theorem~\ref{CN}. Section 4 describes a universal snap-stabilizing algorithm
based upon Mazurkiewicz enumeration algorithm
\cite{MazurEnum}.

\subsection{Related Work}

Given a distributed task, the condition for it being solvable by an
algorithm with explicit termination was first given
in \cite{BVanonymous}.
The presentation we will use in this paper is the one given in
\cite{CGMterm}.
Instead of the
View algorithm of \cite{YKsolvable,BVanonymous},
we use Mazur\-kie\-wicz' algorithm \cite{MazurEnum}. A variation of
Mazur\-kie\-wicz' algorithm was proved to be self-stabilizing in
\cite{selfstabenum}, in the Mazurkiewicz model, a model that offers strong
synchronization between neighbours.  We present here a version for the
cellular model.

Snap-stabilizing algorithms were introduced in \cite{snapstab}. 
A more recent exposition can be found in \cite{recentsnapstab}.
In \cite{breakingin,recentsnapstab}, a general transformation technique is given to
obtain simple snap-stabilizing algorithms from self-stabilizing ones.
The authors expose a snap-stabilizing transformer
for non-anonymous networks which implies that, in networks
with identities, the tasks that are solvable by snap-stabilizing algorithms are exactly the ones that are solvable by self-stabilizing algorithms.
In this paper, we prove the task equivalence between
snap-stabilization and explicit termination in anonymous networks and show that this implies that the expressivity of snap-stabilizing algorithms is different
of self-stabilizing algorithms in the anonymous context.

In \cite{probasnap}, a probabilistic correction condition is proposed
for snap-stabilizing algorithms. A Las Vegas algorithm is an algorithm
whose termination is not guaranteed but whose outputs is always
correct.  The condition of \cite{probasnap} defines,
in a sound way, what is a Las Vegas stabilizing algorithm
that is robust to arbitrary corruption of the
initial memory.

Anonymous networks are networks where nodes do not have a name that is unique.
It has seen many works since the seminal work of Angluin \cite{angluin}. 
There have been two main universal
algorithms proposed to solve problems in this setting.
The first one has been proposed by Yamashita and Kameda in \cite{YKsolvable}. 
Its universality has been extended by Boldi and Vigna in and \cite{BVanonymous} (explicit termination) and \cite{BVselfstab}
(implicit termination). 
It computes 
the (possibly infinite) universal cover of the underlying graph.
The second one computes a minimal base of the underlying graph.
It was presented by Mazurkiewicz \cite{MazurEnum} to solve enumeration.
Its universality has been extended in \cite{GMelection}. Its extension to
numerous other models has been done by Chalopin in \cite{theseJC}, its
application to the Election problem in the message passing model has
been done in \cite{CGMelection}. 
Boldi and Vigna have also shown how to derive a minimal base (in a finite
time) from the universal covering \cite{BVselfstab}. 
One of the main advantage of Mazurkiewicz' algorithm is that it is
always stabilizing, contrary to the View algorithm of
\cite{BVselfstab} where it is necessary to know or derive an estimate
of the size to make it stabilizing.
On the distributed computability side, the first complete
characterization of tasks that admits self-stabilizing algorithms has been 
given in \cite{BVselfstab}. Here, we use a mix of different techniques from the second approach, some of which  were first introduced in \cite{CGMterm}.

There is an unpublished version of Mazurkiewicz'
algorithm in the communication model of this paper but without transient faults
in \cite[chap. 4]{theseJC}, where the model is coined the ``cellular model''.

\section{Definitions and Notations}
\subsection{Basic Definition for Computability}

A network is represented by a graph or digraph $G$
where vertices corresponds to nodes and edges or arcs corresponds to (possibly asymmetric) communication links. The set of vertices is denoted by $V(G)$.
We consider a fixed set of labels $\Lambda$. Labels are used to represent
the local states of parts of the communication network.

So we consider labelled graphs in the general sense. Nodes can be
labelled (internal state of the nodes), arcs can be labelled (messages
in transit, port numbering). We will use $\bG$ to denote a (di)graph with
all its associated labels. Since the input labels can be encoded in the labels, we consider all labelled graphs as the possible inputs for distributed algorithms. The set of all labelled graphs is denoted
$\allg$.
Given a labelled graph $\bG=(G,\lambda)$, where $G$ is  the underlying graph and
$\lambda:V(G)\mapsto\Lambda$ is the labelling function, we will conveniently note
$(\bG,\lambda')$ the graph $G$ labelled by $\lambda\times\lambda'$.

Given a
network $\bG\in\allg$ and a vertex $v$ in $\bG$, we assume that the state
of a node $v$ during the execution of any algorithm is of the form 
$(\lambda(v),\mem(v), \out(v))$. This tuple of registers has the following semantics. $\lambda(v)$ is a read-only part of the state,
$\mem(v)$ is the internal memory of $v$, $\out(v)$ will contain the output
value, i.e. the result of the computation at node $v$.
When the register $\out$ is not defined, it contains the value $\bot$.

A distributed algorithm is an algorithm that is replicated on every
node and operates on the local state of the node $v$ by way of
communication with the neighbours of $v$.
The communication here is done in the \emph{locally shared variables} model
of Dijsktra,
that is also called the \emph{cellular} model \cite{theseJC}. 
A distributed algorithm is a set of rules (pairs of precondition and
command) that describe how a node has to change its current state
(the command) according to its own state and
the state of \emph{all its neighbors} (the
precondition or guard). We say that a rule $R$ is activable at a
node $v$ if the neighborhood of $v$ satisfies the precondition of
$R$. In this case, the vertex $v$ is also said to be activable. If a rule
$R$ is activable in $v$, an atomic move for $v$ consists of reading the
states of all its neighbors, computing a new value of its state
according to the command of $R$, and writing this value to the
register \mem and/or \out. 
If more than one rule is activable at a node, one is chosen non-deterministically.
Of course, it is possible to have priorities for rules, and to discard this non-determinism.

A daemon is a distributed adversary that chooses at each step
a set of activated nodes among the activable ones.
If only one node can be chosen at a time, this is called the \emph{central daemon}.
If any set can occur, this is
called the \emph{asynchronous daemon}.
If the sets of activated nodes is exactly the set of activable nodes
this is called the \emph{synchronous daemon}. 
Given a daemon, an execution, or run, is a sequence of atomic moves of
activated nodes. We consider here the asynchronous daemon
(whose executions contain the synchronous daemon execution).

A vertex-relabelling relation is a relation between labelled graphs
where the underlying graphs are identical.  The evolution of the
global system can be seen as a sequence of relabelling steps where
only the state part of the labels of the graphs is modified, according to
the application of rules prescribed by the algorithm at a set of
locations that depends of the kind of daemon that is considered.
Under a given execution $\rho$, the evolution of the
global configuration of the network \bG %
is described by the sequence of labelled graphs
$\bG,(\bG,\mem_1\times\out_1),(\bG,\mem_2\times\out_2),\cdots$; this is usually
abbreviated to $\bG_0,\bG_1,\bG_2,\cdots$ for convenience.

If the sequence is
finite, that is if there is a step $t\in\N$ where no rule is applicable, or if there
is an infinite suffix starting from step $t\in\N$
where the registers \out are not modified,
we say that the execution has stabilized and denote by $\bG^f$ the 
graph labelled with $\out_t$, $\bG^f=(\bG,\out_t)$.
It is the \emph{terminal state} of the computation.

A terminating problem is a distributed problem for which it is
expected that the nodes have final values. For example, the Election
problem is a terminating problem that should be compared with
the Mutual Exclusion problem where nodes have to solve indefinitely
the problem of entering the critical section one node at a time. We
formally define now what is a terminating distributed problem.
\begin{definition}
  A \emph{terminating task} is a couple $(S,\gfam)$ where
  $\gfam\subset\allg$ is a family of labelled graphs and $S$ is a
  vertex-relabelling relation on \allg.
\end{definition}

The \emph{specification} $S$ is a general way to describe our
distributed problem in terms of relation between inputs and outputs.
This description is independent of the \emph{domain} $\gfam$ where we want to
solve our problem.

For example, the well-known Election problem is specified by $S_{LE}$
such that $\bG S_{LE} \bG'$ if $\bG'=(\bG,\lambda')$ has only one node labelled by the special label \textsc{Elected}. The Size problem where the algorithm
has to compute the number of nodes of the network is specified by
$S_{size}$ such that $\bG S_{size} (\bG,|V(\bG)|)$.

\begin{definition}
  Given a terminating task $(S,\gfam)$,
  an algorithm \algo solves $S$ on $\bG\in\gfam$ if for any
  execution $(\bG_0,\bG_1,\bG_2,\cdots)$ with $\bG_0=\bG$:
  \begin{description}
  \item[decision]
    $\forall v\in V(\bG), \out(v)$ is written exactly once by $v$;
  \item[stabilization]
    the execution stabilizes and 
    the terminal state is denoted $\bG^f$;
  \item[correction]
    $\bG S \bG^f$.
  \end{description}
\end{definition}

\begin{definition}
  The terminating task $(S,\gfam)$ is solvable
  if there exists an algorithm \algo
  such that \algo solves $S$ for all $\bG\in\gfam.$
\end{definition}

When the stabilization is obtained with only finite executions, we say
the algorithm is silent.
When, besides \textbf{correction}, the \textbf{stabilization} property is the only property, we talk
about \emph{implicit termination} (or message termination \cite{Tel}).
When we have both \textbf{stabilization} and \textbf{decision}, we talk about
\emph{explicit termination} (or process termination \cite{Tel}).
In the context of this paper solvability is meant in the explicit termination setting.
Implicit termination is weaker than explicit termination, and for obvious reason, it is the termination for self-stabilizing algorithms.
Note that, in a distant area of Distributed Computing, this is also the termination type of \emph{failure detectors} \cite{ChandraToueg}.
Those are the two main termination mode that are classically considered in
distributed algorithms. See also \cite{CGMterm,GMTterm} for other
types of termination.

\subsection{Self- and Snap-Stabilization}
Informally, a distributed algorithm is said to be self-stabilizing if
an execution starting from any arbitrary global state has a suffix
belonging to the set of legitimate states.  Note that when we consider
the terminating task $(S,\gfam)$,
the set of legitimate states corresponds simply to
the set of $S-$admissible outputs for the given input graph, that is
the set $\{(\bG,\mem,\out)\in\allg \mid \bG\in\gfam, \bG S(\bG,\out)\}$.
So, in the
context of terminating tasks, this corresponds to the definition of
solvability with implicit termination if we require the domain \gfam
to be closed by arbitrary corruption of the initial memory.

More
formally, it is possible to define self-stabilization in the framework
of the previous section. Given a family \gfam, we define
$\overline{\gfam}=\{(\bG,mem)\mid \bG\in\gfam, mem:V(\bG)\to\Lambda\}$.  The
terminating task $(S,\gfam)$ is solvable with self-stabi\-li\-zation if
$(S,\overline{\gfam})$ is solvable with implicit termination.

\medskip
Here we focus on snap-stabilization and give only a formal definition
for snap-stabilization.  Snap-stabilizing algorithms were introduced
in \cite{snapstab}.  A more recent exposition can be found in
\cite{recentsnapstab}.  A snap-stabilizing algorithm computes tasks
that are initiated by ''requests'' at some nodes of the network.  A
request is a special event.  This event is an event exterior to the
algorithm and occurs \emph{after} the end of the faults that led to
arbitrary incorrect values.  Given that the initial memory can be
arbitrarily corrupted, the safety requirement of the problem
specification has to have a special form that takes into account the
fact that starting nodes have seen a request, see
\cite{recentsnapstab}.
In order to have a unified framework, we chose
in our equivalent presentation, to accept any specification $S$ but to
''implement'' the special form in the definition, independently of the
specific specification.

So since the initial memory can be
arbitrarily corrupted, the correction of the \out register is only 
required to be satisfied by nodes that have been causally
influenced by the initial
requests, i.e. nodes for which there exists a sequence of atomic moves
that follow a path originating from a node where a request has been
made.
In other words, a distributed algorithm is snap-stabilizing if
an execution starting from any arbitrary global state has \emph{all
  its causal suffixes} belonging to the set of legitimate states.

Given a specific daemon and an algorithm, the system evolves according
to the daemon and the algorithm: at one step, some nodes are activable
and activated (their actions are processed).
Given an execution $\rho$ on \bG,
that is a sequence $\bG_0,\bG_1,\bG_2...$ of relabelling of \bG where $\bG_0=\bG$,
we denote $A_1,A_2,\cdots$ the sequence of activated nodes.
We have that $\bG_i$ is obtained by applying to $\bG_{i-1}$ the actions for
the nodes of $A_i$.

We proceed to the formal definition.
One or more external actions, \emph{the requests}, are applied at some
nodes $U\subset V(\bG)$.
At time $t$, a node $v$ is \emph{causally influenced} by $U$
if there exists a path $u_0,u_1,\cdots,u_k$ such that $u_0\in U$,
$u_k=v$, and there exists a strictly
increasing function $\sigma:\N\longrightarrow\N$ 
$\forall i\geq 1, u_i\in A_{\sigma(i)}$, and $\sigma(k)\leq t$.

\begin{definition}
  Given a terminating task $(S,\gfam)$,
  an algorithm \algo is snap-stabilizing to
  $S$ on $\bG\in\gfam$ if for any request applied to $U\subset V(\bG)$, 
  \begin{description}
  \item[causal decision]
    $\forall v\in V(\bG)$, $\out(v)$ is written
    exactly once after $v$ has been causally influenced by $U$;
  \item[stabilization]
    the execution stabilizes and 
    the terminal state is denoted $\bG^f$;
  \item[correction]
    $\bG S \bG^f$.
  \end{description}
\end{definition}

\begin{definition}
  The terminating task $(S,\gfam)$ is solvable by snap-stabilization
  if there exists an algorithm \algo 
  such that \algo is snap-stabilizing to $S$ for all
  $\bG\in\overline{\gfam}$. 
\end{definition}

For the sake of simplicity, in the following we always assume that
$\overline{\gfam}=\gfam$.

\subsection{Examples}

To illustrate the various definitions we present in Fig.~\ref{LCR} an
Election algorithm inspired by the well-known Le Lann Chang-Roberts
algorithm \cite{lelann,changroberts}.  We will show that it is
(non-silently) self-stabilizing to the Election task on unidirectional rings,
but that it is not snap-stabilizing.

We consider a unidirectional ring of known size $N$.
The predecessor of a node $v$ is denoted $pred(v)$. Each node $v$ is equipped with a unique identity denoted $id(v)$.
The algorithm maintains two variables $min$ and $ttl$.

\begin{figure}
\begin{rrule}{LCR}
\ritem{Initiate\label{initLCR}}{
  \item $min(v_0)<min(v)$, 
  \item $min(v_0)<id(v_0)$, 
  }{
  \item $min(v_0):=id(v_0)$,
  \item $ttl(v_0):=N$
  }

\ritem{Circulate\label{flood}}{
  \item $min(v_0)>min(v)$,
  }{
  \item $min(v_0):=min(v)$
  \item $ttl(v_0):=ttl(v)-1$
  }

\ritem{Cleaning\label{clean}}{
  \item $min(v_0)\neq min(v)$ or $ttl(v_0)\neq ttl(v)-1$
  \item $ttl(v_0)\neq N$
  }{
  \item $min(v_0):=id(v_0)$,
  \item $ttl(v_0):=N$
  }

\ritem{Election\label{elect}}{
  \item $id(v_0)=min(v)$,
  \item $ttl(v)=1$,
  }{
  \item $\out(v_0)=$\textsc{Elected}
  \item $ttl(v_0)=0$,
  }    
\end{rrule}
\caption{\label{LCR} A LCR Election algorithm. The center of the cell
  is denoted $v_0$, $v$ is $pred(v_0)$.}
\end{figure}

By considering the sequences of consecutive nodes, it is immediate to see that
the labels are stable if and only if the sequence starts from a local minimum and the variables follow the semantic of the propagation of this local minimum according to the original LCR algorithm.
This algorithm is therefore self-stabilizing
but not snap-stabilizing even when adding a special Initiate rule
to deal with the requests as below.

\begin{rrule}{snapLCR}
\ritem{Initiate\label{init}}{
  \item $Request(v_0)$
  }{
  \item $min(v_0)=id(v_0)$,
  \item $ttl(v_0)=N$
  }
\end{rrule}

Indeed any node corrupted in such a way that the Election rule is immediately applicable will incorrectly set its output value to \textsc{ELected} if its predecessor is requested.

\section{Computability of Terminating Tasks}
We start by considering snap-stabilizing terminating tasks. 
We show how the general techniques from explicitly
terminating non-stabilizing tasks can be extended to the
snap-stabilizing case as well.

\subsection{Digraphs and Fibrations}

\subsubsection{Definitions}
In the following, we give the definitions for the tools introduced by
Boldi and Vigna, and extensively studied in \cite{BVfibrations}, to
characterize self-stabilizing tasks in \cite{BVselfstab}.
To introduce the main tool, that is \emph{fibrations}, we need to
consider directed graphs (or digraphs) with multiple arcs and
self-loops. 
A \emph{digraph} $D=(V(D),A(D))$ is defined by a set $V(D)$ of
vertices and a set $A(D)\subset V(D)\times V(D)$ of arcs. Given an arc $a$,  we 
denote $s(a)$ and $t(a)$, the source and target
of the arc.
An undirected graph $G$ corresponds to the digraph $Dir(G)$ obtained
by replacing all edges of $G$ by the two corresponding arcs. In the
following, we will not distinguish $G$ and $Dir(G)$ when the context
permits. 
The family of all digraphs with multiple arcs and
self-loops is denoted $\mathcal D$.
Note that the simple symmetric graphs of $\allg$ have direct counterparts
in $\mathcal D$ via $Dir$.

A dipath $\pi$ of length $p$ from $u$ to $v$ in $D$ is a sequence of
arcs $a_1,a_2,\cdots,a_p$ such that $s(a_1)=u, t(a_p)=v$ and for all
$i$, $s(a_{i+1})=t(a_i)$.
A digraph is strongly connected if there is a path between all pairs
of vertices.
We assume all digraphs to be strongly connected.

Labelled
digraphs will be designated by bold letters like $\bD$, $\bG$, $\bH$
...

A \emph{homomorphism} $\gamma$ between the digraphs $D$ and $D'$ is a mapping
$\gamma: V(D) \cup A(D)\longrightarrow V(D') \cup A(D')$ such that the
image of a vertex is a vertex, the image of an arc is an arc and for
each arc $a\in A(D)$, $\gamma(s(a))=s(\gamma(a))$ and
$\gamma(t(a))=t(\gamma(a))$.
A {homomorphism} $\gamma: V(D) \cup A(D)\longrightarrow V(D') \cup
A(D')$ is an \emph{isomorphism} if $\gamma$ is bijective.

As previously we consider labelled graphs and digraphs. We extend the
definition of homomorphisms to labelled digraphs by adding the
condition they also preserve the labelling
($\lambda(v)=\lambda(\gamma(v))$ for any vertex $v$).

In a digraph $\bG$, given $v_0\in V(\bG)$ and $r\in\N$, 
we denote by $B^\unaryminus_\bG(v_0,r)$ the in-ball of center $v_0$ and radius $r$, that is the
set of vertices $v$ and arcs $a$ 
such that there is a dipath of length at most $r$ from $v$  
to $v_0$.

\subsubsection{Fibrations and Quasi-Fibrations}

The notions of fibrations and quasi-fibrations
enable to describe exactly the ''similarity'' between two
anonymous networks that yields ''similar'' execution for any algorithm
in the model of this paper. For the model of Angluin (used by Mazurkiewicz),
the notions of coverings and quasi-coverings  are the graph morphisms to be used,
see eg. \cite{godard_characterization_2002}.

A digraph $\bD'$ is a fibration of a digraph $\bD$ via $\phi$ if $\phi$
is a homomorphism from $\bD'$ to $\bD$ such that for each arc $a\in A(\bD)$ 
and for each vertex $v\in\phi^{-1}(t(a))$
(resp. $v\in\phi^{-1}(s(a))$),
there exists a unique arc $a'\in\phi^{-1}(a)$ such
that $t(a')=v$ (resp. $s(a')=v$).

The following lemma shows the importance of fibrations when we deal
with anonymous networks. This is the counterpart of the lifting lemma
that Angluin gives for coverings of simple graphs \cite{angluin} and the
proof can be found in \cite{BVelection,BVselfstab,CMelection}.

\begin{lemma}[Lifting Lemma \cite{BVelection}]
\label{lifting} 
If $\bD'$ is a fibration of $\bD$ via $\phi$, then for any daemon, any
execution $\rho$ of an algorithm \algo on $\bD$ can be lifted up to an
execution $\rho'$ of \algo on $\bD'$, such that at any step,
for all $v\in V(\bD')$,
$(\mem(v),\out(v))=(\mem({\phi(v)}),\out({\phi(v)})$.

In particular, when the execution $\rho$ has stabilized, the execution
$\rho'$ has also stabilized and the computed values are the same for
$v$ and $\phi(v)$.
\end{lemma}

In the following, one also needs to express similarity between two
digraphs up to a certain distance. 
The notion of quasi-coverings was
introduced as a formal tool in \cite{MMW,GMelection} for this purpose
in the Mazurkiewicz model. 
The next definition is an
adaptation of this tool to fibrations.

\begin{definition}
Given digraphs $\bK$ and $\bH$, and integer $r$ and $v\in V(\bK)$ and
an homomorphism  $\gamma$ from $B^{-}_\bK(v,r)$ to \bH, \bK is a
quasi-fibration of \bH of center $v$ and radius $r$ via $\gamma$ if
there exists a finite or infinite digraph \bG such that \bG is a
fibration of \bH via a homomorphism $\phi$ and there exists
$w\in V(\bG)$ and an isomorphism $\delta$
from $B^\unaryminus_\bK(v,r)$ to $B^\unaryminus_\bG(w,r)$ such that for any 
$x\in V(B_\bK^-(v,r))\cup A(B_\bK^-(v,r)), \gamma(x) = \phi(\delta(x))$
\end{definition}

If a digraph \bG is a fibration of \bH,
then for any $v \in V(\bG)$ and for any $r\in\N,$ \bG is a
quasi-fibration of \bH , of center $v$ and of radius $r$. 
Conversely, if \bK is a quasi-fibration of \bH 
of radius $r$ strictly greater than the diameter of
\bK, then \bK is a fibration of \bH. 
The following lemma is the counterpart of the
lifting lemma for quasi-fibrations.
\begin{lemma}[Quasi-Lifting Lemma, \cite{CGMterm,CGMelection}]
\label{quasilifting}
Consider a digraph \bK that is a quasi-fibration of \bH of center $v$ 
and of radius $r$ via $\gamma$. 
For any algorithm \algo,
any
execution $\rho$ of \algo on $\bH$ can be lifted up to an
execution $\rho'$ of \algo on $\bK$, such that at any step $t\leq r$,
for all $v\in V(\bK)$,
$(\mem(v),\out(v))=(\mem({\phi(v)}),\out({\phi(v)})$.

In particular, when the execution $\rho$ has stabilized in less than
$r$ steps, the execution $\rho'$ has also stabilized and the computed
values are the same for $v$ and $\phi(v)$.
\end{lemma}

\subsection{Main Result}

In this section we state our main result in Theorem~\ref{CN}. By comparing its statement to that of \cite{CGMterm} we obtain Theorem~\ref{snapequiv}.
It is obvious that the impossibility result of \cite{CGMterm} 
applies here, as well as its proof.
We present the impossibility result integrally here to make the paper
self-contained.

We recall some technical notations and definitions
from \cite{CGMterm}.
\macro{\allv}{\mathcal D_\bullet}
We denote $\allv$ the set 
$\{(\bG,v) \mid \bG\in\mathcal D, v\in V(\bG)\}.$
Given a family $\gfam\subset\allg$, we denote by $\gfam_\bullet$ the set 
$\{(\bG,v) \mid \bG\in\gfam, v\in V(\bG)\}.$
A function $f:\mathcal D\longrightarrow \Lambda\cup\{\bot\}$ is an 
\emph{output function} for a task $(S,\gfam)$ if for each network
$\bG\in\gfam$ the labelling obtained by applying $f$ on each node
$v\in V(\bG)$ satisfies the specification $S$. That is
$\bG S (\bG,\lambda)$ where $\forall v\in V(\bG),$ $\lambda(v)=f(\bG,v)$. 
 
In order to give our characterization, we need to formalize the
following idea.  When the in-ball at distance $k$ of two processes
$v_1$, $v_2$ in two digraphs 
$\bD_1, \bD_2$ cannot be distinguished (this is captured by
the notion of quasi-fibrations and Lemma~\ref{quasilifting}),
and $v_1$ computes its
final value in $k$ rounds, then $v_2$ computes the same final
value.%

\begin{definition}
Given a function $r : \allv \longrightarrow \N\cup\{\infty\}$
and a function $f : \allv \longrightarrow \Lambda\cup\{\bot\}$, 
the function $f$ is $r-$lifting closed if for all 
$\bK,\bH \in \mathcal D$ such that
\bK is a quasi-fibration of \bH, 
of center $v \in V (\bK)$ and of radius $k\in\N$ via the homomorphism
$\gamma$, if $k \geq \min\{r(\bK,v), r(\bH, \gamma(v))\}$, 
then $f(\bK,v) = f(\bH, \gamma(v))$.
\end{definition}

Intuitively, a function $f$ is $r-lifting$ closed if $f(\bG,v)$ depends only
of $B_\bG^-(v,r(\bG,v))$, and it is undefined if $r(\bG,v)=\infty$.

We give now the characterization of terminating snap-stabilizing
tasks. We give the proof of the necessary condition. The converse will
be proved in the following section, by describing a snap-stabilizing
version of Mazurkiewicz' algorithm.

\begin{theorem}\label{CN}
A terminating task $(S,\gfam)$ is solvable by 
snap-stabilization if and only if there exists a function
 $r : \allv \longrightarrow \N\cup\{\infty\}$
and an output function $f : \allv \longrightarrow \Lambda\cup\{\bot\}$
for $(S,\gfam)$ such that,
\begin{theoenum}
\item \label{iterm}%
  for all $(\bG,v)\in\allv$, $r(\bG,v)\neq\infty$ if and only if
  $f(\bG,v)\neq\bot$;
\item \label{rlifting}%
  $f$ and $r$ are $r-lifting$-closed;
\end{theoenum}
\end{theorem}
\begin{qedproof}[[of the necessary condition]]
Consider \algo a distributed algorithm that snap-stabilizes to $S$ on
$\gfam$ in $t$ rounds.

We construct $r$ and $f$ by considering a subset of the possible
executions of \algo.  We consider the synchronous execution of \algo
on any digraph $\bG\in\mathcal D$.  For any $v \in V (\bG)$, if
$\out(v) = \bot$ during the whole execution, then we set $f(\bG, v) =
\bot$ and $r(\bG, v) = \infty$. This is possible since it could be
that $\gfam\varsubsetneq\mathcal D$ and \algo might be not terminating
on graphs not in \gfam.
Let $r_v$ be the first causal step after which
$\out(v)\neq\bot$; in this case, if $r_v\leq t$,
we set $f (\bG, v) = \out(v)$ and $r(\bG, v) = r_v$.
If $t<r_v$, then  we set $f(\bG, v) = \bot$ and
$r(\bG, v) = \infty$.
By construction, \ref{iterm} is satisfied. %

We also show  that $f$ is an output
function and that $f$ and $r$ satisfy 
\ref{rlifting}.  
Consider two digraphs \bK and \bH such that
\bK is a quasi-fibration of \bH, of center $v_0 \in V(\bK)$
and of radius $k$ via $\gamma$ with 
$k \geq r_0 = \min\{r(\bK, v_0 ), r(\bH,  \gamma(v_0 ))\}$.  
If $r_0 = \infty$, then $r(\bK, v_0 ) = r(\bH, \gamma(v_0)) = \infty$
and $f(\bK, v_0 ) = f(\bH, \gamma(v_0)) = \bot.$

Otherwise, from Lemma~\ref{quasilifting},
we know that after $r_0$ rounds, $\out(v_0) = \out(\gamma(v0))$. 
Thus $r_0 = r(\bK, v_0) = r(\bH,\gamma(v0))$ and 
$f(\bK, v_0) = f(\bH , \gamma(v_0)).$ 
Consequently, $f$ and $r$ are $r-$lifting closed.
\end{qedproof}

The previous proof shows that the output function $f$ can be seen as
corresponding to
the final values obtained from the deterministic execution of an
algorithm solving $(S,\gfam)$ under the synchronous daemon.  The
value of $r(\bG, v)$ can be understood as the number of steps needed
by $v$ to compute its final value in \bG.

\section{Main Algorithm}

In this section, in order to obtain our sufficient condition,
we present a general algorithm $\mathcal M_{f,r}$ in
Figure~\ref{mazurSSP} for which we use
parameters that depend on functions $f$ and $r$ corresponding, via
Theorem~\ref{CN}, 
to the terminating task $(S,\gfam)$ we are interested in solving.
This algorithm is a combination of a snap-stabilizing enumeration
algorithm, adapted from \cite{selfstabenum} and a generalization of an algorithm of 
Szymanski, Shy and Prywes (the SSP algorithm for short) \cite{SSP}.

The algorithm in \cite{selfstabenum} is described in a different
model, where each computation step involves some strong
synchronization between adjacent processes.  It is a self-stabilizing
adaptation of an enumeration algorithm presented by Mazurkiewicz in
\cite{mazur}.  The SSP algorithm enables to detect the global
termination of an algorithm provided the processes know a bound on the
diameter of the graph.
The Mazurkiewicz-like algorithm always
stabilizes on any network \bG and during its execution, each process
$v$ can compute an integer $n(v)$ and
reconstruct at some computation step $i$ a digraph $\bG_i(v)$
such that \bG is a quasi-fibration of $\bG_i(v)$ and the image of $v$ is $n(v)$.

By applying the
output function $f$ on $\bG_i(v)$ for $n(v)$,
$v$ can compute its \out value.
However, the enumeration algorithm does not enable $v$ to compute
effectively the radius of this quasi-fibration.  We use a
generalization of the SSP algorithm to compute a counter that is a
lower bound on this radius, as it has already been done in
Mazurkiewicz’ model \cite{GMTterm} and in the message passing model
\cite{CGMterm}. When the SSP counter is greater than $r(\bG_i(v),n(v))$,
the condition on $f$ and $r$ from Theorem~\ref{CN} implies than the
\out value at $v$ is correctly computed for $S$.

\subsection{Modifying Mazurkiewicz’ Enumeration Algorithm}
An enumeration algorithm on a network \bG is a distributed algorithm
such that the \out value are integers and
the result of any computation is a labelling of the vertices
that is a bijection from $V(\bG)$ to $\{1, 2, \cdots, |V (\bG)|\}$.
In particular, an enumeration of the vertices where vertices know
whether the algorithm has terminated solves the Election Problem.  
Since Election is not solvable in all networks, it is not possible to solve the Enumeration problem on all networks.
However, even if not solving Enumeration, in any network \bG,
the Enumeration algorithm of Mazurkiewicz
always stabilizes and yields a digraph $\bG_i(v)$
such that \bG is a quasi-fibration of $\bG_i(v)$.

We give first a general description
of the Mazurkiewicz algorithm.
Every vertex attempts to get its own name in \N\footnote{
this name shall be an integer between $1$ and $|V (\bG)|$ to have an actual Enumeration algorithm. Here we would
need more work to enforce this, however since this is not needed for
our purpose, these technicalities will be skipped. See
\cite{selfstabenum} for a way to get a real Enumeration.}. 
A vertex chooses a name and
broadcasts it together with the name of its adjacent vertices all over the network.  If
a vertex $u$ discovers the existence of another vertex $v$ with the same
name, then it compares its \emph{local view}, i.e., the labelled in-ball of
center $u$ and radius $1$, with the \emph{local view} of its rival $v$. 
If the local view of $v$
is “stronger”, then $u$ chooses another name. Node $u$ also chooses another
name if its appears twice in the view of some other vertex as a result
of a corrupted initial state. 
Each new name is broadcast again over
the network. At the end of the computation it is not guaranteed that
every node has a unique name, unless the graph is fibration minimal. 
However, all nodes with the same name will have the same
local view, i.e., isomorphic labelled neighborhoods.

The crucial property of the algorithm
is based on a total order on local views such that the “strength” of
the local view of any vertex cannot decrease during the
computation. To describe the local view we use the following notation:
if $v$ has degree $d$ and its in-neighbors have names 
$n_1 , \cdots , n_d$ , with
$n_1>\cdots>n_d$ , then $\overline N(v)$, the local view, is the $d-$tuple 
$(n_1 , \cdots , n_d)$.  
Let $T$ be the set of such ordered tuples.  The lexicographic order
defines a total order, $\prec$ , on $T$.  
Vertices $v$ are labelled by triples of
the form $(n, \overline N , M)$ representing during the computation: 
\begin{compactitem}
\item $n(v)\in\N$ is the name of the vertex $v$, 
\item $\overline N (v)\in T$ is the latest view of $v$, 
\item $M(v)\subset\N \times T$ 
  is the mailbox of $v$ and contains all information received at
this step of the computation.  
\end{compactitem}

We introduce other notations. We want to
count the number of times a given name appear in a local view.  For a
local view $\overline N$ , and $n\in \N$, we define $\delta_{\overline N}(n)$ 
to be the cardinality of $n$ in the tuple $\overline N.$
 For a given view $\overline N$ , we denote by $sub(\overline N, n, n' )$ the
copy of $\overline N$ where any occurrence of $n$ is replaced by $n'$.

The complete algorithm is given in Fig.~\ref{mazurSSP}.  The rules are
  given in the \textit{priority order} and $v_0$ denotes the center of the cell (ie the in-ball of radius 1).
\macro{\compN}{\overline{N}} %
\macro{\Pred}{\mathbb P} %

\begin{figure}[t]
\begin{multicols}{2}
\begin{rrule}{Enum}
  \ritem{Initialization\label{label1}}{
    \item $Request(v_0)$
  }{
  \item $n(v_0):=0$,
  \item $\compN(v_0):=N(v_0)$,
  \item $M(v_0):=\emptyset,$
  \item $a(v_0):= -1.$
  }

  \ritem{Diffusion rule\label{dif_rule}}{
  \item There exists $v \in B(v_0)$ such that $M(v)\neq M(v_0)$.
  \item[or] $(n(v_0), N(v_0) ) \notin M(v_0)$,
  \item[or] $\compN(v_0) \neq N(v_0).$
  }{
  \item $M(v_0):=\mathop{\bigcup}\limits_{w\in B(v_0)}M(w)\cup\{(n(v_0),N(v_0))\}$.
  \item $\compN(v_0) := N(v_0).$
  \item $a(v_0):= -1.$
  }

  \ritem{Renaming rule\label{relab_rule}}{
  \item For all $v\in B(v_0), M(v)=M(v_0)$.
  \item $(n(v_0)=0)$ or $( n(v_0)>0
    \mbox{ and there exists }(n(v_0),N)\in M(v_0)
    \mbox{ such that }((N(v_0)\prec N)))$.
  \item $n(v_0) > 0 \text{ and } \exists (n_1 , N_1 ) \in M(v_0 )$
    such that $\delta_{N_1}(n(v_0))\geq 2.$
  }{
  \item $n(v_0) = 1+\max \{n\in\N\mid (l,n,N)\in
    M(v_0)\,\,\text{for some}\,\, l,N \}$.
  \item  $M(v_0) = M(v_0)\cup
    \{(n(w),N(w)) | w \in B(v_0)\}$,
  \item $a(v_0)=-1$.
  }
  \ritem[gSSPfix]{Fix gSSP counter}{
    \item If there exists $v\in B(v_0), \; |a(v) - a(v_0)| \geq 2$ or $(M(v)\neq M(v_0)$ and $a(v_0)\neq-1)$
  }{
  \item  $a(v_0) := -1.$
  }
  \ritem[gSSP]{gSSP rule}{
    \item $\forall v\in B(v_0), \; M(v) = M(v_0),$ $|a(v) - a(v_0)| \leq 1$ and $\neg\Pred(v_0)$
      }{
    \item  $a(v_0) := 1 +  \min\{a(v) \mid v\in B(v_0)\}.$
      }
  \ritem[Decision]{Output rule\label{Mdecision}}{
    \item For all $v\in B(v_0), \; M(v) = M(v_0)$ and $\Pred(v_0)$
      }{
    \item  $\out(v_0) = f(\bK(v_0),w(v_0))$
      }

\end{rrule}
\end{multicols}
\caption{\label{mazurSSP}Snap-stabilizing algorithm
  $\mathcal M_{f,r}$.  The parameters are the functions $f$ and $r$
  from Theorem~\ref{CN}.
    $\bK$ is defined by a
  local procedure and the predicate \Pred depends on $r$.
}
\end{figure}

\macro{\strong}{\textsc{Strong}}

The labeling function obtained at the end of a run
$\rho$ of Mazurkiewicz’ algorithm is noted $\pi_\rho$. 
If $v$ is a vertex of \bG, the couple $\pi_\rho(v)$ associated with $v$ is
denoted $(n_\rho(v), M_\rho(v)).$ 
We also note the final local view of $v$ by $N_\rho(v).$ 
For a given mailbox $M$ and
a given $n \in \N,$ we note $\strong_M(n)$ the local view 
that dominates all $\compN, (n,\compN) \in M$ (\ie $\compN \prec \strong_M (n).$
Except for the first corrupted stages, $\strong_{M(v)}(n)$ is actually the
“strongest local view” of $n.$

\begin{theorem}
A run $\rho$ of Mazurkiewicz’ Enumeration
Algorithm on \bG with any initial
values finishes and computes a final labeling $\pi_\rho$
verifying the following conditions for all vertices $v, v'$
of $V(\bG)$~:
\begin{theoenum}
\item $M_\rho(v) = M_\rho(v').$
\item $\strong_{M_\rho(v')}(n_\rho(v)) = \compN(v) = N_\rho(v).$
\item \label{injlabel} $n_\rho(v) = n_\rho(v')$ if and only if $N_\rho(v) = N_\rho(v')$.
\end{theoenum}
\end{theorem}
\begin{qedproof}
  Even if the model is different, beside technicalities,
  this can be proved similarly
  to the proof of \cite{selfstabenum}. %
\end{qedproof}

Now we explain how it is possible to extract the map of a minimal
base. This is usually done by considering the graphs induced by the
numbers and associated local views that have maximal views. However,
here, due to the arbitrary initial failures, the mailbox should be
cleaned up before use. It is possible to have some maximal $(n,\compN)$ but
$n$ does not actually exists on any $v$.

Finally, each vertex shall compute locally the set of
actual final names from the final mailbox $M_\rho$. 
We note $\bG_\rho$ the graph defined by
\begin{eqnarray*}
V_\rho &=& \{n_\rho(v) | v \in V(\bG)\},\\ 
A_\rho &=& \{(n_\rho(v_1 ), n_\rho (v_2)) | (v_1 , v_2 ) \in A(G)\}.
\end{eqnarray*}

For a mailbox $M$ and an integer $n$, we define the
set $V^M(n)$ by induction.
\begin{eqnarray*}
V^M_0 &=& \{n\},\\
V^M_{i+1} &=& V^M_i \cup\{ t | \exists s \ V^M_i , \delta_{\strong_M(s)}(t)= 1 \}.
\end{eqnarray*}

If $i_0$ is such that $V^M_{i_0} = V^M_{i_0+1}$ then we define 
$V^M(n) = V^M_{i_0}$. Finally, we have,
\begin{lemma}[\cite{selfstabenum}]
For all $v\in V(\bG)$, $V^{M_\rho} (n_\rho (v)) = V_\rho$.
\end{lemma}

By defining $A^M$ by $\{(n_1,n_2)|n_1,n_2\in V^M(n)\mbox{ and
}\delta_{\strong_M(n_1)}(n_2)=1\}$, we obtain a graph
$\bG_{M(v)}=(V^{M(v)},A^{M(v)})$.  We can not readily use $\bG_{M(v)}$
since it could be that it is not in \gfam.  We denote by $\bK(v)$ a
digraph that is in \gfam and that is a quasi-fibration of $\bG_{M(v)}$
of radius $a(v)$ and of center $w(v)$.  Such a digraph can be found by
a local procedure enumerating all graphs and vertices of $\gfam_\bullet$ until
one is found. This semi-algorithm will always terminate because of the
following property.

\begin{proposition}
  \label{quasifib-base}
  Let $P$ be the set of requesting processes. Let $v$ that has been
  causally influenced by $P$, and such that $a(v)\geq 0$. 
  The graph \bG is a quasi-fibration of
  $\bG_{M(v)}$ of center $v$ and radius $a(v)$.
\end{proposition}
\begin{qedproof}
  We add that every $w\in B(v,a(v))$ has been influenced to the
  statement and prove this new statement by induction on $i$, the
  number of steps since $P$ has received the requests.

  Initially, at step 1, the requests are being processed by Enum1, \ie the set
  of influenced nodes is $P$ and the property holds trivially.

  Assume the property holds at step $i$ and consider $v_0$ a vertex
  that is activated at round $i+1$.  We have to consider two cases,
  either $v_0$ was already influenced at round $i$ or it is a newly
  influenced node.

  If $v_0$ is a newly influenced node. The only rule of interest is
  gSSP because other rules are setting $a(v_0)$ to $-1$.  But we show
  that $v_0$ cannot apply this rule. Indeed, assume
  $M(v_0)\neq\emptyset$, then, the causality path to $v_0$ starts in a
  root whose variables have been reset, and from which the causality
  chain of applications will propagate its new name. So $M(v_0)$ has
  to be updated to, at least, this name before being able to apply
  gSSP.

  If $v_0$ has already been influenced then the induction statement
  applies at the previous round. Denote $a(v_0)$ the value of the
  counter at the end of round $i$ and assume that for all $v\in
  N(v_0), a(v)=a(v_0)$. We prove that the statement
  holds for $a(v_0)+1$ at round $i+1$.

  If $a(v_0)=0$ then, by the same argument as in the previous case, 
  the neighbours of $v_0$ have all been influenced and the statement
  holds with a radius $1$.

  If $a(v_0)>0$ then the neighbours have been influenced by induction
  assumption. Moreover, every $v\in N(v_0)$ is the center of a
  quasi-fibration of radius $a(v_0)$. Therefore, $v_0$ is the center
  of a quasi-fibration of radius $a(v_0)+1$.
  Similarly, every $w\in B(v,a(v_0))$ has been influenced and
  the ball $B(v_0,a(v_0)+1)$ is totally influenced.
  The statement holds at round $i+1$.
\end{qedproof}

The algorithm from Fig.~\ref{mazurSSP} uses the functions $f$ and $r$ given in the necessary condition of Theorem~\ref{CN}.
The two functions are used to define a digraph \bK (defined above) and
a predicate \Pred defined below.
The predicate needs to make the counter $a$ to increase when what can be extracted
from the mailboxes (that is the minimum base of \bG)) is the same locally.
But it must also make the algorithm stop when there is enough
information to conclude. This information is enough when
the value $r$ for the reconstructed base matches the counter of stability $a$.

\begin{theorem}
With $\Pred(v) := (a(v) < r(\bK(v),n(v)))$, the algorithm 
$\mathcal M_{f,r}$ snap-stabilizes to $S$ for any set $P$ of requested nodes.   
\end{theorem}

\begin{qedproof}
Consider a node $v$ just after it has applied rule \ref{Mdecision},
we have $\strong(M(v))$ that is constant in the neighbourhood,
$r(\bK(v),n(v))\leq a(v)$ 
and $out(v)=f(\bK(v),w(v)).$
Since, by construction, 
$\bK(v)$ is a quasi-fibration of $\bG_{M(v)}$ of radius 
$a(v)\geq r(\bK(v),n(v))$ and of center $n(v)$, and since $f$ and $r$
are $r-$lifting closed, $\out(v) = f(\bK(v),w(v)) =
f(\bG_{M(v)},n(v)),$
and $r(\bK(v),w(v)) = r(\bG_{M(v)},n(v)).$
 From Prop.~\ref{quasifib-base}, since $a(v) \geq r(\bG_{M(v)},n(v)$
 and since $f$ is $r-$lifting closed, $\out(v) =
 f(\bG_M(v),n(v))=f(\bG,v).$

 Since $f$ is an output function for $(S,\gfam)$, the \out labels are correct for $S$ in \bG.
\end{qedproof}

\subsection{Complexity}
The algorithm $\mathcal M_{f,r}$ is a universal algorithm and
therefore for given $s,\gfam)$ it can have a bigger complexity than a
tailored algorithm.  However it should be noted that the complexity of
$\mathcal M_{f,r}$ is divided in two components, the stabilization of
the Enumeration part and the increase of the SSP counter until it is
greater than $r$. Note that the former depends on the graph \bG only
and that the latter depends on the family \gfam.  The complexity from
the Enumeration has been shown in \cite{selfstabenum} to be, in the
Angluin model, at most $t |V(\bG)|^2$ where $t$ is the sum of the number
of vertices and of the highest name $n$ initially known. The proof can
be extended to the model of this paper.

\section{Conclusion}
We have shown that for anonymous networks, the terminating tasks that
can be solved by a snap-stabilizing algorithms are exactly the ones
that can be solved by a distributed algorithm with explicit
termination. This complements the already known task-equivalence
between self-stabilizing terminating tasks and distributed tasks
computed with implicit termination.  
The important consequence is that the partial knowledge (like bound on
the size, diameter etc ...) that could be used to get explicit
termination in the non-stabilizing case are also the ones that can be
used to have snap-stabilizing solutions.

A limit of this result is that it does not give the intrinsic complexity of a problem and it could be that solving a problem by snap-stabilization is harder than solving it with explicit termination. The computability is equivalent however whether the complexity is also equivalent is an open problem.

For lack of space, we do not discuss probabilistic snap-stabilization
\cite{probasnap}. It is not difficult to see that 
the techniques presented here enable to prove that a
task has a probabilistic snap-stabilizing solution if and only it has
a (non-stabilizing) Las Vegas solution.

An interesting open question, as in the
self-stabilizing case, would be to find a \textit{direct} way to
transform any given 
anonymous algorithm into a snap-stabilizing one.
Such transformation might have benefits regarding the complexity.

\medskip
The author wishes to thank Jérémie Chalopin for sharing ideas and fruitful
discussions about distributed computability in various settings,
including some closely related to this paper.

\input{snapstab.bbl}

\end{document}

%% file: snapstab.bbl
\newcommand{\etalchar}[1]{$^{#1}$}